\documentclass[12pt]{article}
\usepackage{amssymb}
\usepackage[graph,curve]{xy}

\newtheorem{theorem}{Theorem}
\title{Generating connected acyclic digraphs uniformly at random}
\author{Guy Melan\c con, Fabrice Philippe\\
LIRMM CNRS UMR 5506\\
Montpellier, France\\
{\tt \{Guy.Melancon, Fabrice.Philippe\}@lirmm.fr}}
\date{\today}
\begin{document}
\maketitle \abstract{ We describe a simple algorithm based on a Markov 
chain process to generate simply connected acyclic directed graphs 
over a fixed set of vertices.  This algorithm is an extension of a 
previous one, designed to generate acyclic digraphs, non necessarily 
connected.  }


\bibliographystyle{plain}
\noindent 
\section{Introduction}
Combinatorial objects can sometimes be investigated through random 
generation in order to discover various properties, estimate 
statistics or test conjectures.  Also, objects might be generated 
randomly to test the behaviour or performance of algorithms, with the 
hope that all possible or interesting cases will be obtained from the 
generation process.  For example, graph drawing algorithms can be 
tested or studied by running them over a large set of examples.  This 
approach is widely used and many authors provide test sets on the web 
(see \cite{steinlib, patrignani} for instance).

Graphs are widely used by non-mathematicians as a modelling tool in 
various fields (social sciences, computer science and biology to name 
just a few).  A common approach is to develop a theoretical but 
specific framework based on graphs and to prove its relevancy and 
usefulness by experimentally testing it on many examples.  However, 
the lack of "real-world" example sets motivate the construction of 
artificial ones.  Randomly generating graphs appears as a solution, 
provided that the generating process can be somewhat controlled or 
kept within a given class of graphs.

Acyclic digraphs (directed graphs containing no circuits) prove to be 
appropriate in many situations.  The absence of circuits can indeed 
have different interpretations (logical ordering of tasks or 
processes, ordering induced from time, etc.).  Many authors have 
studied their properties providing insight on the global and average 
shape of acyclic digraphs \cite{robinson_labeled, robinson_unlabeled, 
bender1986, gessel, mckay}.  An algorithm generating acyclic digraphs 
uniformly at random has only been published recently 
\cite{eurocomb01}.  However, the authors did not provide any specific 
conditions showing how to use their algorithms to generate {\em simply 
connected} acyclic digraphs.  This condition is often seen as being 
central in many applications and motivated our work.  As we shall see, 
a careful investigation of the Markov chain introduced in 
\cite{eurocomb01} reveals how and why the same algorithm can be 
slightly modified and used to generate simply connected acyclic 
digraphs uniformly at random.

This note is organized as follows.  We first recall basic results and 
notations and describe the Markov chain introduced in 
\cite{eurocomb01}.  We then discuss the conditions under which it can 
be adapted as to provide an algorithm for generating simply connected 
acyclic digraphs.  The last part of the note is devoted to the proof 
of our result.

\section{A Markov chain algorithm}
Let $N \geq 2$ be a fixed integer and let $V = \{1$, \ldots, $N\}$ 
denote a finite set of {\em vertices}.  We consider the set ${\cal A}$ 
of all acyclic directed graphs over $V$, that is, graphs containing no 
circuits\footnote{Recall that a circuit in a directed graph $G$ is a 
sequence of vertices $v_0$, $v_1$, \ldots, $v_k = v_0$ where each 
$(v_i, v_{i+1})$, $i = 0\ldots k-1$, is an arc in $G$.}.  Next, we 
define the Markov chain $M$ over the set ${\cal A}$.  Each state of 
$M$ is an acyclic digraph in ${\cal A}$.  Because the set $V$ of 
vertices is fixed, we will not distinguish between a digraph in ${\cal 
A}$ and the set of its arcs.  The transition between any two states in 
$M$ is given as follows.  Let $X_t$ denote the state of the Markov 
chain at time $t$.  Suppose a couple of integers $(i, j)$ has been 
drawn uniformly at random from the set $V \times V$.
\begin{itemize}
\item[($T_1$)]
If $(i, j)$ is an arc in $X_t$, it is deleted from $X_t$. That is,  
$X_{t+1} = X_t \setminus (i,j)$.
\item[($T_2$)]
If $(i, j)$ is not an arc in $X_t$, then 
\begin{itemize}
\item[(i)] it is added to $X_t$, provided that the resulting graph is acyclic. 
That is, $X_{t+1} = X_t \cup (i,j)$.
\item[(ii)] otherwise, nothing is done. That is, $X_{t+1} = X_t$.
\end{itemize}
\end{itemize}

The probability of a transition going from a state $X$ to a state 
$Y\neq X$ is either 0 or $1/N^2$.  Note that the transition rules make 
the transition matrix $A$ symmetric.  This is a key property in 
proving that the Markov chain converges towards the uniform 
distribution.  More precisely, we have:
\begin{equation}
{\label{eq:limit}}
\lim_{n \to \infty} A^n \cdot d = \bar 1,
\end{equation}
where $d$ is any initial distribution over ${\cal A}$ and $\bar 1$ 
stands for the normalized vector $(\frac{1}{|\cal A|}$, \ldots, 
$\frac{1}{|\cal A|})$.  The other key element in the proof is that the 
state space of $M$ is {\em irreducible}.  That is, given any two 
acyclic digraphs $G_1$ and $G_2$, there exists a sequence of 
transitions going from $G_1$ to $G_2$.  This is straightforward since, 
starting from $G_1$, one can travel by rule $(T_1)$ to the graph with 
no arc (which is obviously acyclic) then by rule $(T_2)$ to $G_2$.

This result can be converted into an algorithm for randomly generating 
acyclic digraphs, each having the same probability of being drawn.  
Indeed, condition $(T_2)$ relies on the ability to test whether the 
addition of $(i, j)$ would create a circuit, which can be easily done 
\cite{rodeh_alon}.  Hence, starting from a graph with an empty set of 
arcs one can iteratively apply the rules $(T_1)$ and $(T_2)$ to build 
an acyclic digraph.  Applying those rules a large (but finite) number 
of times leads to an algorithm, which randomly outputs acyclic digraphs 
(with an {\em almost} uniform distribution).

\section{Customizing the algorithm}
In \cite[Theorem\ 1]{eurocomb01}, the authors prove that random 
acyclic digraphs have $N^2/4$ arcs on average.  However, in many 
applications, the total number of edges in graphs is proportional to 
their number of vertices\footnote{When generating graphs to test 
drawing algorithms, for instance, it makes no sense to generate graphs 
with too many edges.  Graphs having $4N$ edges are actually considered 
as {\em dense} in the Graph Drawing community.}.  Hopefully, the chain 
can be adapted to these situations by imposing further conditions on 
the transition ($T_2$).  For instance, one can allow the insertion of 
a new arc provided that, additionally to condition ($T_2$), the total 
number of arcs in the graph does not exceed a given upper bound.  The 
condition can also be refined in order to impose an upper bound on 
each vertex individually.

Two observations must be made here.  First, it is important to mention 
that additional conditions imposed on transitions $(T_1)$ or $(T_2)$ 
actually induce a restriction on the state space on which they act.  
The corresponding transition matrix may not be symmetric any more, so 
that other sufficient properties have to be established.  Also, the 
irreducibility of this space has to be established before we can 
conclude that Eq.\ \ref{eq:limit} holds for the restricted transition 
matrix.  However, the situations considered in \cite{eurocomb01} are 
rather simple and in each case the irreducibility of the restricted 
space is straightforward.

The contribution of this short note is to show how the Markov chain 
can be adapted to produce simply connected acyclic 
digraphs\footnote{Recall that a directed graph $G$ is {\em simply 
connected} if its underlying non directed graph is connected.  That 
is, if for any two vertices $u, v$ there exists a sequence of vertices 
$u = v_0$, $v_1$, \ldots, $v_p = v$ such that either $(v_i, v_{i+1})$ 
or $(v_{i+1}, v_i)$ is an arc in $G$ ($i = 0\ldots p-1$).}.  In the 
sequel, we will refer to simply connected graph as connected graph, 
for sake of brevity.

In many applications, this appears to be an essential property in the 
modelling phase.  Indeed, social networks are rarely made of several 
disconnected components, since it is exactly the interactions between 
the actors of the network that motivates their study.  Also, 
disconnected acyclic digraphs are irrelevant in modelling task 
synchronization problems, since tasks belonging to distinct components 
can be performed independently.  As a further example, chronology 
forces acyclicity in Bayesian networks.

An obvious approach one can adopt in order to adapt the Markov chain 
$M$ to generate connected acyclic digraphs is to impose on the rule 
$(T_1)$ an additional condition: the deletion of an arc should only be 
allowed if the arc is not \emph{disconnecting}, that is, if the 
resulting graph is still connected.  Furthermore, a disconnecting arc 
belongs to no (undirected) cycle, so that it may be reversed without 
creating a circuit.  This seemingly superfluous rule has two 
advantages at least: it enables further transitions, slightly 
shortening the diameter of the transition graph, and it makes the 
chain irreducible for $N=2$.  Observe that the modified conditions 
preserve the symmetry of the transition matrix, admitted that the 
state space underlying this restricted Markov chain is now the set of 
all {\em connected} acyclic digraphs.  So, for instance, the graph 
with no arc does not belong to the state space anymore.  The minimal 
number of arcs is now $N-1$.  Each of these connected digraphs with 
$N-1$ arcs can actually be built by constructing a (non directed) tree 
and by assigning appropriate orientations to its edges.

For sake of clarity, we denote by ${\cal A}'$ the set of all connected 
acyclic digraphs and modify the transition rules as follows.  Take 
uniformly at random $(i,j)$ in $V^2$.
\begin{itemize}
\item[($T'_1$)]
If $(i, j)$ is an arc in $X_t$ then 
\begin{itemize}
\item[(i)] if it is not disconnecting, it is deleted from $X_t$. 
That is,  $X_{t+1} = X_t \setminus (i,j)$.
\item[(ii)] if it is disconnecting, it is reversed. 
That is,  $X_{t+1} = X_t \setminus (i,j)\cup (j,i)$. 
\end{itemize}
\item[($T_2$)]
If $(i, j)$ is not an arc in $X_t$, then 
\begin{itemize}
\item[(i)] it is added to $X_t$, provided that the resulting graph is acyclic. 
That is, $X_{t+1} = X_t \cup (i,j)$.
\item[(ii)] otherwise, nothing is done. That is, $X_{t+1} = X_t$.
\end{itemize}
\end{itemize}
Write $M'$ for this Markov chain running under the rules $(T'_1)$ and 
$(T_2)$.  Since the transition matrix is symmetric, the problem we 
face is to show that the state space ${\cal A}'$ underlying $M'$ is 
irreducible.  Observe that, contrarily to what was done in 
\cite{eurocomb01}, this is not straightforward.  Indeed, it is not 
immediate to see that given two {connected} acyclic digraphs $G_1$ and 
$G_2$ there exists a sequence of transitions from $G_1$ to $G_2$ and 
going through connected acyclic digraphs only.  The proof is given in 
the next section.

In \cite{eurocomb01} the authors mention an algorithm by Alon and 
Rodeh \cite{rodeh_alon} to test for circuits in directed graphs which 
can be used for conditions $(T_2)$.  Any algorithm testing a directed 
graph for simple connectedness can be used for condition $(T'_1)$.  
This can be performed in a time proportional to the number of arcs in 
the graph (using a depth first search for instance, see, e.g., the 
textbook by Jungnickel \cite{jungnickel}).

The complexity of the algorithm is dominated by the total number of 
iterations.  The problem of estimating a good upper bound of this number is a
difficult one and is at the
centre of the theory on Markov chains \cite{sinclair}.  As we will see in the 
next section, the transition graph is compact, with diameter in the 
order of $N^2$.

\section{Irreducibility of the Markov chain}
We shall now give the complete proof of the irreducibility of the 
Markov chain $M'$.  It is worthwhile to point out that the result 
holds, with the same proof, if $\cal{A}'$ is replaced by any subclass 
containing all digraphs on $V$ with $N-1$ and $N$ arcs.

\begin{theorem}
{\label{thm:main}} Let $M'$ be the Markov chain defined over the space 
${\cal A}'$ of all connected acyclic digraphs, together with the 
transitions defined by the rules $(T'_1)$ and $(T_2)$.  The chain $M'$ 
is irreducible, that is, given two connected acyclic digraphs $G$ and 
$H$, there exists in $M'$ a sequence of transitions $G = G_0 \to G_1$, 
$G_1 \to G_2$, \ldots, $G_{p-1} \to G_p = H$ $(p \geq 1)$ with all 
$G_i \in {\cal A}'$ $(i = 0\ldots p)$.  Moreover, such a sequence 
exists with length at most $(N+7)(N-3/2)$.
\end{theorem}

{\em Proof of Theorem\ \ref{thm:main}}\\
Since transitions in $M'$ are reversible, it suffices to show that 
there exists a sequence of transitions from any given connected 
acyclic digraph $G$ to a fixed element in $\cal{A}'$.  We proceed in 
three steps.  Let $H$ be any acyclic digraph over $V$, denote by $\bar 
H$ the undirected graph obtained from $H$ by ignoring the orientations 
of its arcs.

\emph{Step 1} There exists in $M'$ a sequence of transitions from $G$ 
to a directed tree $T$ (i.e., an acyclic digraph such that $\bar T$ 
contains no cycle).  Indeed, $\bar T$ can be obtained by computing a 
spanning tree for $\bar G$.  The transition is then defined by a 
sequence of arc deletions by rule $(T'_1)$, leaving only the arcs 
present in the spanning tree.

\begin{figure}[h]
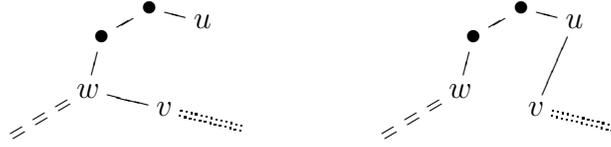

$$
\xygraph{ 
!{0;(.8,-,2):0}
!~:{@{==}}
{w} ( 
:[ld(0.8)],
-[r]{v} !~:{@{::}}:[r],
-[u(0.7)]{\bullet}-[ur(0.5)]{\bullet}-[r(0.7)]{u}
)}
\qquad\qquad\xygraph{ 
!{0;(.8,-,2):0}
!~:{@{==}}
{w} ( 
:[ld(0.8)],
[r]{v} !~:{@{::}}:[r],
-[u(0.7)]{\bullet}-[ur(0.5)]{\bullet}-[r(0.7)]{u}-"v"
)}
$$
  \caption{{\small Turning the leaf $u$ into a vertex of degree 2}}
  \label{star}
\end{figure}

\emph{Step 2} There exists in $M'$ a sequence of transitions from $T$ 
to a directed chain $C$ (i.e., a directed tree over $V$ such that $\bar C$ is a
chain; for short, a \emph{dichain}).  We proceed by diminishing the number
$\ell(T)$ of the leaves in $\bar T$.  Suppose  
$\ell(T)>2$.  Starting from a leaf $u$, one can find in $\bar T$ a 
simple path $u= u_1,u_2, \ldots,u_p=w$ such that $u_i$ has degree 2 
for all $i = 2, \ldots,p-1$ and $w$ has degree at least 3. Let $v\neq 
u_2$ be another neighbour of $w$ (see the left-hand part of Figure\ 
\ref{star}, where the dashed lines stand for one or more edges and the 
dotted ones for possible edges).
By rule $(T_2)$, either the arc $(v, u)$ or $(u, v)$ can be added to 
$T$ without creating a circuit, after what the arc 
between $w$ and $v$ cannot be disconnecting.  Deleting it by 
rule $(T'_1)$, we get a directed tree 
$T'$ such that $\ell(T') =\ell(T) - 1$, because $w$ has still degree at 
least 2 and $u$ is not a leaf anymore.  The result follows by 
iterating until $\ell(T)=2$. 

\begin{figure}[h]
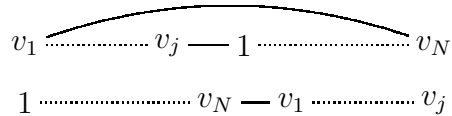

$$
\xygraph{ 
!~:{@{..}}
{v_1}:[r(1.5)]{v_j}-[r(0.8)]{1}:[r(2)]{v_N}!~:{@{-}}:@/_15pt/ "v_1", 
!~:{@{..}}[d(0.6)]{1}:[r(2)]{v_N}-[r(0.8)]{v_1}:[r(1.5)]{v_j}
}
$$
 \caption{{\small Turning the vertex 1 into a leaf}}
  \label{chain}
\end{figure}

\begin{figure}[h]
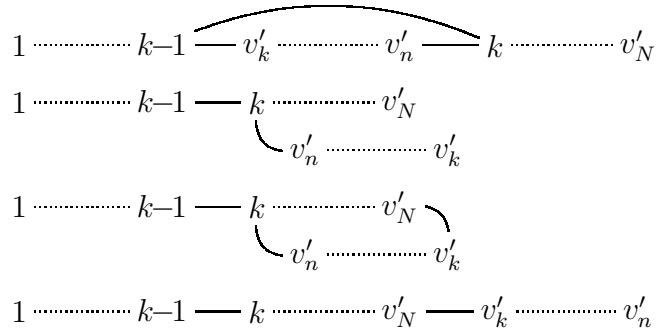

$$
\xygraph{ 
!~:{@{..}}
{1}:[r(1.5)]{k\!\! -\!\! 1}="x"-[r]{v'_k}:[r(1,5)]{v'_n}-[r]{k}
(:[r(1.5)]{v'_N},!~:{@{-}}:@/_15pt/ "x"), 
!~:{@{..}}
[d(0.6)]{1}:[r(1.5)]{k\!\! -\!\! 1}-[r]{k}
(:[r(1.5)]{v'_N},!~:{@{-}}:@/_9pt/[d(0.5)r(0.5)]{v'_n}!~:{@{..}}:[r(1,5)]{v'_k}),
[d(1.7)]{1}:[r(1.5)]{k\!\! -\!\! 1}-[r]{k}
(:[r(1.5)]{v'_N}="xx",!~:{@{-}}:@/_9pt/[d(0.5)r(0.5)]{v'_n}
!~:{@{..}}:[r(1,5)]{v'_k}!~:{@{-}}:@/_9pt/"xx"),
!~:{@{..}}
[d(2.8)]{1}:[r(1.5)]{k\!\! -\!\! 1}-[r]{k}:[r(1.5)]{v'_N}-[r]{v'_k}:[r(1,5)]{v'_n}
}
$$
 \caption{{\small Placing the vertex $k$ (orientations are omitted).}}
  \label{chain2}
\end{figure}

\emph{Step 3} There exists in $M'$ a sequence of transitions from $C$ 
to any Hamiltonian directed chain, say $\{(i, i+1), 1\leq i<N\}$.  By rule
$(T'_1)$ indeed, this holds for $N = 2$ 
trivially.  Set $N\geq 3$.  By rule $(T'_1)$, it suffices to show 
that there is in $M'$ a sequence of transitions from $C$ to one of 
the dichains associated with the chain $\{\{i,i+1\},1\leq i<N\}$.  Let 
$\bar{C}=\{\{v_i, v_{i+1}\}, 1\leq i<N\}$, we first bring $C$ into a 
dichain $C'$ such that $\bar{C'}=\{\{v'_i, v'_{i+1}\}, 1\leq i<N\}$ 
and $v'_1=1$.  If there is $j>1$ such that $v_j=1$, an arc 
between $v_N$ and $v_1$ may be added to $C$ by rule $(T_2)$, and the 
arc in $C$ relative to the edge $\{v_{j-1},v_j\}$ may then be deleted by 
rule $(T'_1)$ (see Figure \ref{chain}, where 
orientations of the arcs have been omitted for sake of simplicity, and vertices
joined by dotted paths are possibly identical).
Now $v'_1=1$ in $C'$, let $k$ be the least integer such 
that $v'_k\neq k$.  Suppose $k<N$, then the (simple) path 
in $\bar{C'}$ from 1 to $k$ is longer than the one from 1 to $v'_k$.  
We next bring $C'$ to a dichain $C''$ such that $\bar{C''}=\{\{v''_i, 
v''_{i+1}\}, 1\leq i<N\}$ with $v''_i=i$ for $1\leq i\leq k$.  Since 
its justification is quite similar to the previous one, we only 
describe the construction in Figure \ref{chain2}.  Iterating it until $k=N$ 
completes the proof of the third step. 

Finally, let $G_1$ and $G_2$ be in $\cal{A}'$.  Then $G_i$ 
$(i=1,2)$ may be turned into a directed tree $T_i$ by $N(N-1)/2-(N-1)$ 
transitions at most, the worst case obtaining when $G_i$ is a 
tournament.  Then $T_i$ may be turned into a dichain $C_i$ by 
$2((N-1)-2)$ transitions at most, the worst case obtaining when $T_i$ 
is a star.  Finally, $C_1$ may be made a Hamiltonian dichain by 
$(N-1)/2$ arc reversions at most, and $C_2$ may be turned into the 
latter Hamiltonian dichain by $2+4(N-2)$ transitions at most.

\smallskip\noindent\textbf{Remark} For $N\geq 3$, the reversion rule 
$T'_1(ii)$ is not indispensable for Theorem\ \ref{thm:main} to hold.  
It is not clear whether it fastens convergence of the process or not.

\noindent\textbf{Acknowledgements} We wish to thank our colleague 
Alain Jean-Marie for stimulating discussions.  This work originally 
started from email exchanges with Fabio G. Cozman.  
\footnote{\tt{http://www-2.cs.cmu.edu/$\sim$fgcozman/}}

\begin{small}\bibliography{dagsconnexes}\end{small}
\end{document}